# NOVAE EJECTA AS DISCRETE ADIABATICALLY EXPANDING GLOBULES

**Robert Williams**

Space Telescope Science Institute, 3700 San Martin Drive, Baltimore, MD  21218, USA;  wms@stsci.edu

**ABSTRACT**

Available data for novae show that the X-ray and visible spectral regions correlate with each other as they evolve.  Large differences in ionization exist simultaneously in the two wavelength regimes, and a straightforward model is proposed that explains the characteristics observed in both spectral regimes.  Its key features are (1) ejected blobs of very high density gas from the white dwarf that expand to create within each clump a wide range of emitting density, ionization, and velocity, and (2) a more homogeneous circumbinary envelope of gas that is produced by secondary star mass loss. The relative mass loss rates from the two stars determine whether the He/N or the Fe II visible spectrum predominates during decline, when hard X-rays are detected, and when the white dwarf can be detected as a super soft X-ray source.

*Short Title:*  Adiabatically Expanding Nova Blobs

## 1.  INTRODUCTION

The nova phenomenon involves the major components of mass transfer binary systems:  the white dwarf (WD), the secondary star, and the accretion disk.  For systems that are highly magnetic an additional component exists in the form of accretion columns that funnel gas onto and out of the magnetic poles of the WD.  There are also indications of a fifth component, circumbinary gas, that may exist in some novae systems at outburst.  Spatial resolution of individual binary systems is still beyond present capabilities, even with radio interferometry, so the behavior of the individual components of cataclysmic variables (CVs) must be inferred almost entirely from spectroscopy.  In fact, the present paradigm for CVs owes much of its existence to spectroscopic analyses, e.g., that of Kraft (1964), that demonstrated the existence of WDs and accretion disks in novae and dwarf novae.  The ejecta of a small fraction of novae do become resolvable months after the outburst when they have achieved a sufficient spatial extent, and they typically show strong inhomogeneities

A rich archive of postoutburst optical spectra of novae exists that more recently has been augmented by spectral studies of the X-ray and radio regimes.  A common theme of spectral analyses across different wavelength bands is that separate emitting components exist in order to explain the spectra as they evolve.

## 2.  EVIDENCE FOR MULTIPLE COMPONENTS

The most graphic evidence for the existence of two ejecta components comes from images of the resolved nova shell of Nova GK Persei/1901 (Shara et al. 2012), which show very elongated head-tail structures of ejected condensations moving through an ablating ambient medium.  There is uncertainty whether the ambient medium is interstellar gas or ejecta, e.g., a wind, from the nova outburst.  Shara et al. demonstrate that the morphology of a subset of the emitting clumps argues for their immersion in a wind from GK Per itself, evidence for separate components with different velocities.

Spectroscopic evidence for multiple components of ejecta has also been inferred from the continua of the different energy regimes and from different behavior in the evolution of observed emission and absorption lines. X-ray intensities observed in different passbands by the *ROSAT*, *BeppoSAX*, *Chandra*, and *SWIFT* X-ray satellites have shown many novae to radiate thermal-like continua of temperatures $T\sim10^{5-6}$ K superposed on a flat continuum produced by shocks that extends to higher energies. Clear evidence for multi-component X-ray emission comes from analyses of the novae V1974 Cyg/1992 (Balman, Krautter, & Oegelman 1998), V382 Vel/1999 (Orio et al. 2002; Ness et al. 2005), V2491 Cyg/2008 (Page et al. 2010), V1723 Aql/2010 (Krauss et al. 2011), and the recent Nova Mon 2012 (Ness et al. 2012; Orio & Tofflemire 2012).

Most novae become soft X-ray sources at some point during their decline, exhibiting an approximate thermal component emanating from the hot WD surface, although the source of this 'surface' emission could in many cases actually be ejected globules. Such globules would initially radiate a thermal continuum before colliding with other ejected clumps that then excite the very high ionization X-ray emission lines and the variability observed in the X-ray region (Orio et al. 2002; Page et al. 2010).

The radio emission of postoutburst novae has been observed interferometrically, especially by the (E)VLA, which has characterized the time and spatial evolution of novae and their radio spectra (Hjellming et al. 1979; Taylor et al. 1987; Chomiuk et al. 2012). The large majority of novae emit thermal radiation that may on occasion also be accompanied by synchrotron radiation. For some novae radio emission becomes detectable within 2-3 weeks of outburst, although for others it does not initially appear and increase in strength until many weeks after maximum light, e.g., the recurrent nova T Pyxidis (Nelson et al. 2013).

A standard model for nova radio emission was developed by Hjellming et al. (1979) based on a Hubble-flow model for ejected gas emitting thermal bremsstrahlung, and it has been used to explain the basic spectral evolution of novae (Seaquist & Bode 2008). Recently, spectral and temporal coverage of novae have shown the Hubble-flow model to have limitations in explaining observed radio fluxes (Krauss et al. 2011). There is evidence for large inhomogeneities in the ejecta whose emitting properties differ both in physical parameters and time evolution, and may involve shocks. Shocks are also inferred in novae ejecta from the high energy X-rays that the *SWIFT* satellite frequently observes in novae. The consensus view is that proper modeling of nova radio emission requires multiple, distinct components of gas having widely different physical conditions.

The visible spectra of novae pass through an initial optically thick dense phase with a low temperature continuum and P Cygni line profiles for the stronger lines, evolving to a low density optically thin forbidden line spectrum. Individual novae show large changes in their visible spectra with time although this in itself does not require multiple components to explain the spectral changes. The evolution of ejecta from an optically thick 'fireball' stage that expands until its lower density gives rise to a forbidden emission spectrum can be explained satisfactorily by models of changing mass loss rate, density, and temperature distribution of the WD ejecta (Hauschildt et al. 1997; Shore et al. 2011).

There are indications from the optical spectra of a more complex situation, which has led to an alternative view advocated to explain the origin of the two classes of spectra, He/N and Fe II, that novae exhibit after outburst through visible light maximum (Williams 2012). A consistent picture is advocated where He/N spectra originate from the WD ejecta whereas the Fe II spectra are formed in gas lost from the secondary star. The spectral evolution of the recent outburst of the recurrent nova T Pyx shows behavior that is supportive of two distinct components that produce different spectra in early decline.

The data of Shore et al. (2011) and Ederoclite (2013) show T Pyx to be a hybrid nova that initially radiated a He/N spectrum after outburst which then transitioned into an Fe II spectrum within 7 days. After two months it evolved back into an He/N spectrum that was very similar to its initial pre-maximum spectrum. Five spectra that document this evolution from 15 April—17 June 2011 are shown in Figure 1, taken with the ESO VLT+X-Shooter and displayed over a broader wavelength interval by Ederoclite (2013). One of the most striking features is the similarity of the two He/N spectra, the first and last of the sequence, especially in specific details such as the relatively narrow red component of the N II λ5005 emission feature, which is not present in the Fe II spectral phase.

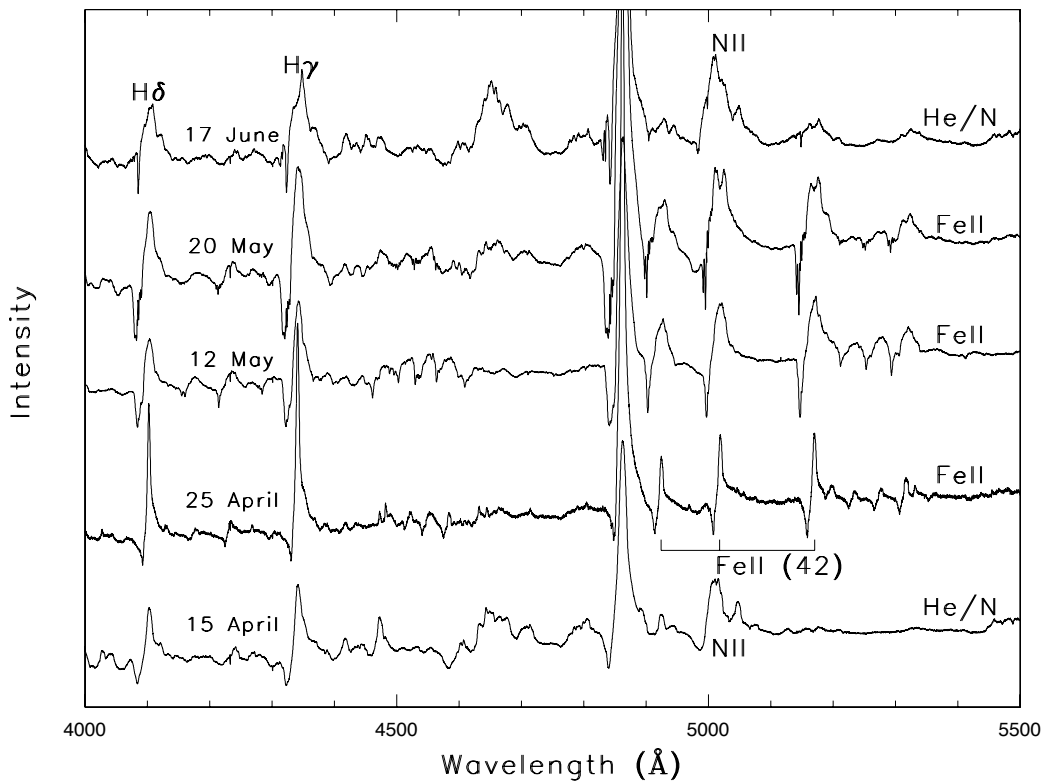

Fig. 1 – Spectral evolution of T Pyxidis in the two months following outburst obtained with VLT+X-Shooter at a spectral resolution of R=17,000. The wavelength region surrounding Hβ includes He I and N II lines and Fe II multiplet 42.

The similarity of the initial April and the June spectrum, both He/N class, argues for their origin in the same component of gas. The fact that the narrower velocity structure of the lines in the middle three spectra, all Fe II class, differ from those of the He/N spectra is indicative that the Fe II spectra originate in a separate, lower expansion velocity component of gas rather than in a single component that has simply changed physical conditions. The 'Fe II' gas must envelope the He/N emitting gas because the He/N spectral features are obscured during the Fe II phase. Since the Fe II spectrum has the same flux level as the He/N spectrum one cannot argue that the Fe II spectrum is simply much brighter than the He/N component. The observations require that the more rapidly expanding He/N emitting gas be confined to a region smaller than and within that of the Fe II gas, which is optically thick in the continuum and obscures the He/N spectrum. Otherwise, the He/N lines should be seen. The Fe II gas subsequently dissipates, allowing the He/N component once again to become dominant. It remained so between the 2-month solar avoidance period from early July to mid-September, at which time the nova spectrum had already evolved into its forbidden nebular spectrum.

The He/N and Fe II classification system referred to above is based upon overall general characteristics of postoutburst spectra that emphasize emission features and physical conditions (Williams 1992). It differs from the long-standing classical description of novae spectra defined by McLaughlin (1960), whose spectral classification system was based largely on the properties of the absorption lines. Some aspects of the two systems can be reconciled.

McLaughlin defined four absorption phases which generally appear in sequence: the pre-maximum, principal, diffuse enhanced, and Orion spectral phases.  The diffuse enhanced spectrum, which occurs around maximum light, corresponds to our Fe II phase, which is the most common spectrum observed for ~80% of classical novae.  Although McLaughlin asserts that it appears after the 'principal' spectral phase, we find that not to be the case.  The 'principal' spectrum defined by McLaughlin corresponds to transient heavy element absorption (*thea*) systems analyzed by Williams et al. (2008), whose sudden appearances with narrow absorption features and lower expansion velocities during the Fe II phase correspond to a distinctly different component of gas than the Fe II diffuse enhanced gas.  McLaughlin's Orion phase, featuring the N III $\lambda$4645 Bowen multiplet, 'nitrogen flaring' and an '[O III] flash' is not one whose spectral characteristics have been observed by this author as a separate spectral phase nor adhering to the chronology McLaughlin has attributed in nova spectral development.  The two different classification systems are thus complementary in some aspects, with the McLaughlin system, based on photographic spectra and being largely descriptive and phenomenological, differing from the 'Tololo' system used above, based on digital data and defined in terms of the physical parameters of the emitting gas.

### 3. X-RAY AND OPTICAL SPECTRA

Few would disagree that analyses of X-ray, radio, and visible spectra all point to the existence of strong inhomogeneities and variability in the emitting and absorbing gas of novae.  The debate is whether the spectra can all be explained entirely by mass loss from the WD alone or whether separate components of gas, viz., from the secondary star, make major contributions to the spectra.  Shore et al. (2013) have presented a solid case for the former that involves a recombination wave that passes through the ejecta.  Our purpose here is to call attention to the fact that a different, self-consistent alternative can be formulated that explains basic features of the spectral evolution that involve components of mass loss from both stars.

Discussions of X-ray observations of classical novae have been presented by Orio, Covington, & Oegelman (2001) for *ROSAT* observations, and by Ness et al. (2007) and Schwarz et al. (2011) for *SWIFT* data.  Moderate spectral resolution observations of novae with the *Chandra* and *XMM-Newton* satellites have frequently shown a strong emission-line spectrum of highly ionized CNO ions, e.g., C VI, N VII, and O VII, superposed on a continuum having multiple components (Orio 2012).  Novae exhibiting such line spectra have generally done so when the optical spectrum is either in the He/N phase or later in the decline during the nebular forbidden emission line phase.  X-rays are weak at best when the visible spectrum is in the Fe II phase.  Also common for novae months after outburst when the expanding ejecta have become less dense and transparent is the super soft source (SSS) phase of X-rays emitted at peak energies E<1 keV by the then visible hot WD, commonly occurring during the nebular phase.

The rapid maneuverability of *SWIFT*, with its low resolution instruments that can distinguish between low- and high-energy X-rays with relatively brief observations, allows multiple targets to be followed daily.  Schwarz et al. (2011) have characterized the X-ray evolution of novae observed with *SWIFT* in the 0.4-10 keV range since it began observations in 2004.  Figure 4 of their paper summarizes the behaviors of sources that have been sufficiently bright to be observed multiple times during postoutburst.  A common theme emerges from the data:  hard X-rays ($\geq$1 keV) are often observed in the days following outburst.  This phase can last for days to months and it generally transitions to lower energies, becoming dominant in an SSS phase as the object fades.  The interpretation of this behavior is that the WD ejecta interact with each other or with circumbinary gas collisionally to produce hard X-rays, and as outburst activity ramps down mass loss decreases and the dissipating ejecta reveal the hot WD surface regions.

The visible spectra of novae during the early hard X-ray phase tend to conform to the He/N spectral class that ~20% of novae exhibit near maximum visible light. The He/N spectral class is dominated by moderate ionization lines, e.g., N II, He I, and He II which are formed at the same time and with similar velocity profiles as the very high ionization X-ray emission lines.

## 4. ADIABATICALLY EXPANDING GLOBULES

The explosive ejection of recently semi-degenerate gas by the TNR on the WD surface will almost certainly produce ejecta that are clumped and inhomogeneous, both from inevitable variations in the temperature that drives the TNR and from Raleigh-Taylor instabilities. The resulting environment is one of adiabatically expanding clumps of gas having a distribution of sizes, each blob of which has a wide range of density and temperature, decreasing outwardly from the center of the blob to the rapidly dissipating outer layers. An obvious model of novae ejecta follows from this situation: initially very high density homogeneous spheres of temperature $T\sim10^{7.5}$ K and number density $n\sim10^{28}$ cm$^{-3}$ are ejected from the WD, as required by the critical pressure $P_{crit}\sim10^{19}$ dyn cm$^{-2}$ required for a surface TNR (Starrfield, Sparks, & Truran 1986; Yaron et al. 2005). Assuming adiabatic expansion with $\gamma = 5/3$, the outer surfaces of the clumps begin expansion at the initial sound speed of $v_{surf}\sim10^3$ km s$^{-1}$. At these high expansion velocities, which are applicable primarily to the outer regions because the inner regions are confined by the overlying gas, T and n decrease rapidly with time. The decreasing outer density drives down the temperature adiabatically as $T \propto n^{\gamma-1}$, subject to possible influence by conduction and radiative heating and cooling processes.

The resulting large temperature gradient within each clump creates a wide range of ionization stages that can account for the low ionization Na I and Fe II lines observed in novae visible spectra so soon after highly energetic outbursts. Given the initial conditions assumed above the gas densities that pertain to temperatures of $T\sim10^4$ K are of order $n\sim10^{22}$ cm$^{-3}$, which (1) exceeds normal photosphere densities, (2) suggests that the Stark effect could be a significant source of line broadening in early novae spectra, and (3) is consistent with the high densities indicated by the relative intensities of the O I $\lambda$7773/$\lambda$8446 lines (Williams 2012). With time the above range of parameters migrates inward (in Lagrangian coordinates) during the lifetime of the globule, which depends on the initial size of the condensation.

For condensations of diameter $r_c<10^3$ m the majority of the gas in the expanding clump will cool to relatively low temperatures of $T<10^3$ K within minutes of ejection, especially since small globules are optically thin and therefore not heated by gamma rays. Collisions of the expanding condensations with each other form a mixture of hot shocked gas surrounding the WD that radiates hard X-rays.

The adiabatic expansion of a homogeneous gas sphere is a classic problem in hydrodynamics. Formulation of the basic equations is straightforward but full analytic solutions are difficult to obtain without approximations (Sedov 1959; Landau & Lifshitz 1987), whereas numerical calculations yield detailed results for the velocity structure, density, and temperature. Results for very high density gas have not yet appeared in the literature, however published results representing hot inhomogeneities in a perfect gas, viz., the earth's upper atmosphere, are available (Dogra et al. 2001; Dogra & Wadsworth 2005) and these calculations have applicability to the blobs of gas ejected from the WD by the outburst.

Hot clumps of upper atmospheric gas can be created by electrical and turbulent activity and they expand into the ambient gas, producing shocks. Dogra and colleagues have used the 'direct simulation Monte Carlo' method (Bird 1994) to follow the temperature, density, and expansion velocity evolution of an initially uniform temperature and density sphere. The profiles of prototypical clump parameters from a sample calculation of an expanding globule are shown in Figs. 2-4, reproduced from Dogra et al. (2001). The plots indicate how conditions vary with time in an adiabatic expansion into a cooler ambient medium.

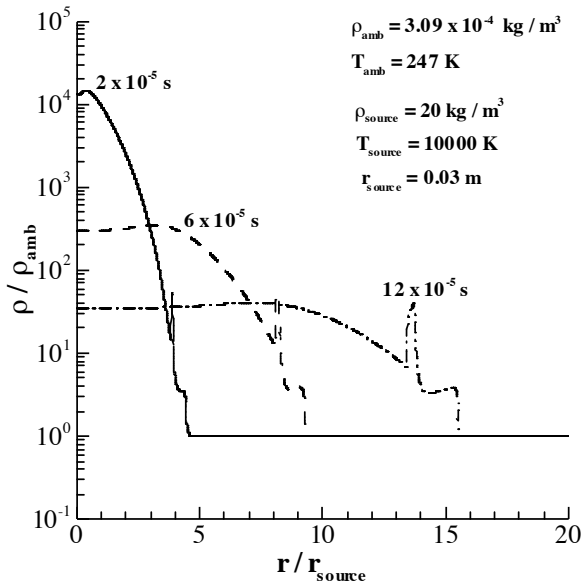

Fig. 2 -- Evolution of density in an adiabatically expanding sphere (from Dogra et al. 2001, with permission of the authors and the AIP)

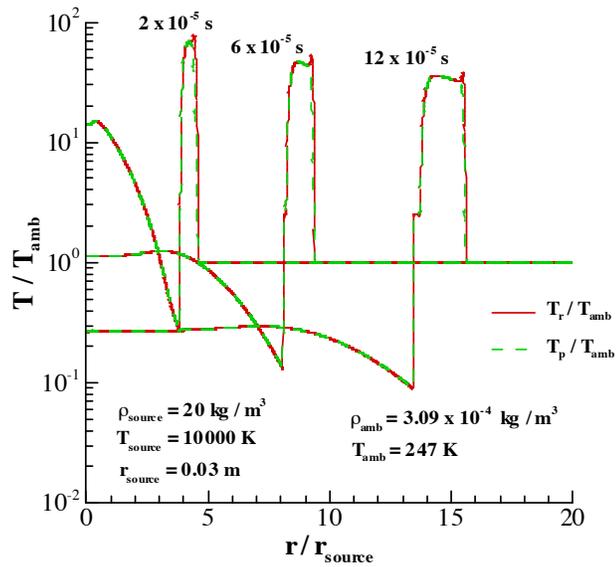

Fig. 3 -- Evolution of temperature in an adiabatically expanding sphere (from Dogra et al. 2001, with permission of the authors and the AIP)

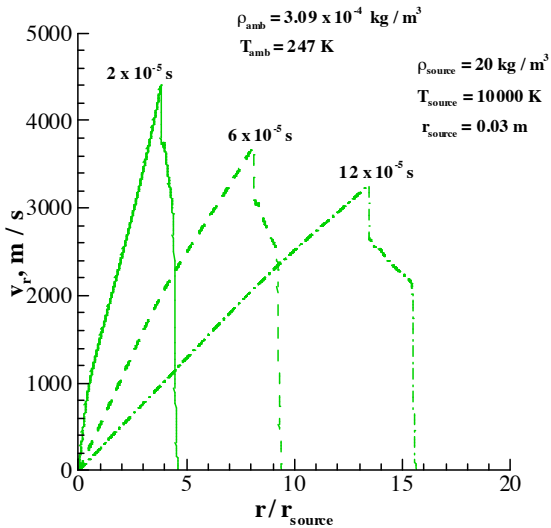

Fig. 4 – Evolution of local expansion velocity in an adiabatically expanding sphere (from Dogra et al. 2001, with permission of the authors and the AIP)

Key characteristics of expanding gas spheres can be inferred from Figs. 2-4. After the initial sound-crossing time for the globule the local expansion velocity at each point increases roughly linearly outward toward the outer surface, dictated by the confining pressure of the overlying gas which causes the inner gas to expand more slowly, keeping it hotter, denser, and more highly ionized. The roughly linear velocity law is a good approximation for spherical adiabatic expansion, referred to as the 'linear velocity approximation' in the Sedov solution for a filled blast wave (Ostriker & McKee 1988).

The expansion velocity at each co-moving point in the globule decreases with time due to the progressively smaller sound speeds that follow from the adiabatically decreasing temperatures. The local temperature is set by its adiabatic dependence upon density except in those outer regions of the clump that experience heating from the shock produced if it expands into an ambient medium rather than a vacuum. The clump steadily dissipates in time with the central density and temperature falling at a rate that is dictated by the expansion and by heating from the reverse shock that propagates back into the clump from the outer surface. The gradients in $T(r)$ and $n(r)$ in the inner regions continually flatten from the confining pressure of the overlying gas as the clump dissipates, and they take on a flat profile radially for the central part of the globule. The inner regions of larger clumps have correspondingly lower expansion velocities by the time they reach the same densities that smaller clumps pass through more quickly. This situation produces a natural setting for a broad range of emitting ions and transitions with a correspondingly broad range of temperatures and critical densities.

## 5. SPECIFIC NOVAE: OPTICAL—X-RAY SPECTRAL CORRELATIONS

The large differences in the gas components that radiate the X-ray and optical spectra present a major challenge for models of the ejecta. Dense clumps of rapidly expanding gas ejected by the outburst can account for the disparate features of novae spectra in the different wavelength bands. Descriptions of postoutburst visible spectra are normally documented in IAU Circulars, The Astronomer's Telegram, and for southern novae on Fred Walter's SMARTS website[1]. Compilations of X-ray spectra observations can be found in Ness et al. (2007) and Schwarz et al. (2011) for *SWIFT* data, and also in the general review by Orio (2012). We have searched the literature to identify a sample of novae for which good coverage has been obtained in both optical and X-ray regions during decline in order to look for correlations between the visible and X-ray spectra. For most novae the data are either too fragmentary or lacking in optical—X-ray simultaneity to provide meaningful constraints, however the following objects do provide useful information:

**V382 Vel/1999:** Early optical spectra of this nova, obtained 5 days after maximum light, showed it to be of the Fe IIb class (Della Valle et al. 2002), and it evolved to a neon-rich nebular forbidden phase within one month of outburst. Mukai & Swank (1999) did not detect X-rays on day $+6^d$, but both Orio & Ricci (1999) and Mukai & Ishida (2001) reported detections on days $+15\text{-}20^d$ with *BeppoSAX* and *ASCA* of an X-ray continuum of temperature $kT \sim 8$ keV observed through an absorbing gas layer of column density $N_H \sim 10^{23}$ cm$^{-2}$ that steadily decreased in subsequent months. These observations were followed seven months later by lower resolution *BeppoSAX* observations that showed the nova to be in a super soft source phase. Nine months after outburst, when the visible spectrum was in the nebular forbidden-line phase Ness et al. (2005) obtained a *Chandra* spectrum that showed strong lines of O VIII, Ne X, Mg XII, etc. in the X-ray region. Notably, no Fe lines were observed in this X-ray spectrum even though Fe XVII and Fe XVIII transitions would normally be expected with the other lines that were observed. Their absence is significant given that Fe II lines were so prominent in the early visible spectra. This is very suggestive that the early Fe II spectrum formed in a different component of gas than the later X-ray emission line spectrum.

**V2362 Cyg/2006:** Munari et al. (2008) observed this nova to be a typical Fe II class object, remaining in the Fe II phase as it slowly transitioned to the nebular spectrum after 10 months. In the intervening time the nova reversed its initial decline in brightness and achieved a secondary brightness peak 8 months after outburst. The nova was initially detected in X-rays with *SWIFT* six months after outburst while the spectrum was still in the Fe II permitted phase, before forbidden lines had appeared (Lynch et al. 2008). Dust formed shortly after the secondary maximum after which the optical forbidden spectrum developed as the novae steadily declined in brightness. Little information is known about the X-ray spectrum other than that it appeared to be harder than that of a normal SSS (Lynch et al. 2008).

---

[1] http://www.astro.sunysb.edu/fwalter/SMARTS/NovaAtlas/atlas.html

**V458 Vul/2007:** This nova declined rapidly following outburst and was unusual in achieving several secondary peaks in visible brightness in the first three weeks following outburst. Poggiani (2008) found the optical spectrum to be a standard Fe II type during this period. One month after outburst, and after the secondary brightness maxima events had occurred, the spectrum transformed into a He/N spectrum, thereby becoming a hybrid. The only subsequent record of an optical spectrum was by Prater et al. (2007), who observed nebular forbidden lines in December 2007, four months after outburst.

No X-ray detections occurred within the first months of the outburst, particularly during the visible Fe II spectral phase. On day $+71^d$ the *SWIFT* satellite began regular X-ray observations of the nova with an initial detection of a hard X-ray continuum during the time the visible spectrum was changing from the He/N phase to the forbidden-line spectrum (Ness et al. 2009). On day $+88^d$ *Suzaku* observed an emission spectrum of Kα lines of O VIII, Ne X, and S XV (Tsujimoto et al. 2009) superposed on a 1 keV continuum. The strong line spectrum persisted for many months until it evolved into a supersoft source ten months after outburst. The optical and X-ray components decayed slowly thereafter.

**V2491 Cyg/2008:** This nova declined very rapidly, with $t_2=5^d$, and was studied spectroscopically by Munari et al. (2011). There is some ambiguity about the early visible spectral class. Tomov et al. (2008) commented on the predominance of numerous strong, broad Fe II multiplets the first week after outburst. However, the spectra displayed by Munari et al. indicate an apparent He/N class at this time. Between days $+9-20^d$ the nova reversed its decline and achieved a secondary maximum. Immediately thereafter, Munari et al. (2011) data showed that the optical spectrum was of the He/N class during the interval $+22-33^d$. The nova would be unique if the spectrum were of the He/N class during the intervening secondary peak because all other novae known to have experienced secondary brightness peaks have radiated Fe II-class spectra (Williams 2012). No further spectra are known for this object until day $+108^d$, when Munari et al. observed the nova spectrum to be nebular with forbidden lines of [Ne III], [O III], [Fe VII], and [Fe X].

X-rays from V2491 Cyg were observed by *SWIFT* and have been described by Page et al. (2010). Faint detections of hard X-rays occurred within a day of the outburst, becoming weaker for the next 10 days after which the nova began optically re-brightening to its secondary peak. A strong increase in X-rays began at this time, continuing until day $+40^d$ as the X-ray energy distribution became progressively softer as the nova evolved into a strong SSS source. This took place as the visible spectrum was evolving from its He/N phase to the nebular spectrum. During this time X-ray flickering was observed on timescales of minutes. After day $+50^d$ the X-ray count steadily decreased for the next 100 days, transitioning from a soft continuum spectrum to that of an optically thin plasma (Page et al. 2010) as it steadily faded from detection.

**V2672 Oph/2009:** This was another extremely fast nova, with $t_3=4^d$, having very broad emission lines indicating expansion velocities in excess of 5,000 km s$^{-1}$ and a large extinction of E(B-V)=1.6 due to its relatively large distance and low galactic latitude. Munari et al. (2011) documented the early visible spectral evolution, noting a definite He/N spectral class and similarity to that of the recurrent nova U Scorpii. F. Walter also obtained spectra that appear on his SMARTS website that confirm the He/N class and the considerable strength of He II λ4686 and λ5412. The nova became too faint to obtain useful spectra after 3 weeks, before any significant change to the initial spectra and before nebular forbidden lines had emerged. Schwarz et al. (2009) activated a ToO program on *SWIFT* that detected faint levels of X-rays for one month beginning with a detection on the first day after outburst. Within the observational uncertainties dictated by the very low counts a soft continuum and a harder flux component were inferred and spectral fits indicated a line of sight absorbing column density of $N_H \sim 10^{22}$ cm$^{-2}$.

**U Sco/2010:** A very fast recurrent novae whose many outbursts have been well studied, U Sco underwent its tenth recorded outburst in 2010 that spectroscopically was very similar to its previous outbursts. The visible spectrum of the nova was followed by numerous groups (Yamanaka et al. 2010; Diaz et al. 2010; Schaefer 2010; Kafka & Williams 2011; Mason et al. 2012; Maxwell et al. 2012) who documented a clear He/N class spectrum, with evidence for the presence of weak Fe II features, that transitioned to a nebular spectrum between days +23-51$^d$ with [Ne III], [N II], and [O III] evident (Diaz et al. 2010). The decline was extremely fast, with $t_3$=3$^d$, consistent with the broad line widths indicating very high expansion velocities in excess of 5,000 km s$^{-1}$, and optical flaring was observed beginning on day +10$^d$.

X-ray observations of U Sco began immediately at the time of outburst. The *SWIFT* X-ray counts increased by more than two orders of magnitude over the first 10 days, and *Chandra* and *XMM-Newton* spectra on days +19$^d$, +23$^d$, and +35$^d$ showed both an SSS component from the hot WD and also X-ray lines from a highly ionized plasma (Ness et al. 2012; Orio et al. 2013). All three spectra showed a continuum of temperature $T_{WD}$=(7.5--9) × 10$^5$ K representing the hot WD surface region, and high ionization represented by optically thin lines from N VII, O VIII, and Ne X. The observations were notable for the detection of eclipses in both the X-ray and the optical spectra (Mason et al. 2012), requiring a significant fraction of the radiation in both the visible and X-ray bands to originate well within the WD Roche lobe. Evidence of a Thomson scattering corona of size larger than the binary orbit was also inferred from the eclipse observations (Ness et al. 2012; Orio et al. 2013). The eclipse observations also provided evidence for the postoutburst re-formation of the accretion disk.

**T Pyx/2011:** The visible spectrum of the most recent outburst of recurrent nova T Pyx was followed at high spectral resolution by Shore et al. (2011, 2013) and Ederoclite et al. (2013), and numerous lower resolution spectra have also been posted on F. Walter's Stony Brook/SMARTS Nova Atlas website. This outburst produced the hybrid spectral evolution discussed in §2 and shown in Fig. 1, alternately exhibiting He/N and Fe II class spectra before transitioning to a typical optical nebular spectrum.

Soft X-rays were detected by *SWIFT* within a day of the outburst and became steadily weaker for the next 10 days. For the period +15-110$^d$ after outburst there were no X-ray detections, during which time the visible spectrum was almost entirely of the Fe II phase, and at the end transitioning to the nebular phase. Beginning on day +112$^d$, at the time the optical spectrum had evolved to the nebular spectrum, and until one year after outburst *SWIFT* began regular observations that documented an increasing X-ray flux that was predominantly soft, characterized by a thermal-like continuum of energy kT=50-80 keV, and also variability on timescales of hours (Kuulkers et al. 2011). A hard component was also present during this period which was shown from *Chandra* X-ray spectra (Tofflemire et al. 2011) to consist of lines of highly ionized CNO and Fe.

The combined spectra of the above objects, representing widely different energy regimes, are crucial for piecing together a workable model for novae ejecta. The SSS phase is almost certainly due to the hot WD surface region when it becomes visible without significant obscuration from a high column density of intervening ejecta, and the high ionization line spectrum is produced by energetic collisions among ejected blobs of gas (Krautter 2002; Orio 2012). The consensus explanation for the visible spectrum, from the different permitted phases to the final low density nebular spectrum, has normally invoked direct mass loss from the WD in the form of a variable wind (Hauschildt et al. 1997; Kato & Hachisu 2011; Shore et al. 2011).

## 6. EXPANDING GLOBULES MODEL

The above objects that have been observed simultaneously at X-ray and optical wavelengths exhibit the following characteristics: (i) there is an epoch immediately after outburst when X-rays are detected at low flux levels. The epoch is brief for novae of Fe II spectral class at maximum light, but is typically much longer for He/N novae. (ii) X-ray emission is generally weak during the optical Fe II spectral phase, e.g., often not detected by *SWIFT*, and it tends to be SSS when detected. (iii) Stronger super soft source phases tend to occur later in decline, during the visible nebular forbidden line phase. (iv) X-ray emission line spectra of high ionization are usually associated with He/N and nebular visible spectral classes, and they do show notable short-term variability over timescales of minutes and hours. (v) U Sco eclipse observations show the X-rays to originate in a small region near the WD, also indicated independently by the minute-scale X-ray flickering of V2491 Cyg.

These characteristics can all be accommodated by a model in which the WD ejects discrete blobs that expand immediately and whose lower density outer layers produce the He/N visible spectrum. High velocity collisions between the globules create hard X-rays with the bulk of this activity occurring relatively near the WD, i.e., within ~10-100 $R_{WD}$. This results in (a) an X-ray emitting region that is variable in size, (b) time variations in flux, and (c) the simultaneous emission of a low ionization spectrum in the visible and a very high ionization X-ray spectrum, both having similar velocity distributions.

Ejected blobs provide a natural explanation for discrete absorption features that are observed in novae, many of which eventually manifest themselves as emission features in the optically thin nebular stage. In the case of HR Del/1967, which like GK Per/1901 shows a multi-clumped shell (Harman & O'Brien 2003), Gallagher & Anderson (1976) were able to correlate the most prominent individual absorption components in high resolution spectra with distinct emission features having the same radial velocities months later in the nebular spectrum.

If an optically thick circumbinary wind or envelope of gas from the secondary star develops to produce the Fe II visible spectrum, as is the case for the majority of novae, the inner activity is obscured by the circumbinary gas which normally completely obscures everything with the WD Roche lobe, leading to weaker X-ray flux. Less X-ray activity is then observable during the Fe II phase but it re-emerges when the circumbinary Fe II gas dissipates to reveal continuing WD ejecta and/or the WD surface regions. In this way one can account for how the more slowly expanding Fe II phase gas surrounds and obscures the more inner confined, more rapidly expanding WD ejecta responsible for hard X-ray emission. There is evidence from highly ionized X-ray line spectra observed during the optical Fe II phase that some of the denser, presumably larger, high velocity ejecta blobs do tunnel through the Fe II gas to become visible, and these blobs are the source of both visible and X-rays.

Several comments should be made about the above paradigm. First, the relatively homogeneous Fe II circumbinary shell constitutes a legitimate expanding pseudo-photosphere, whereas the discrete blobs that produce the He/N spectrum may not always because they do not necessarily have a large covering factor. Second, the WD almost certainly ejects a hail of pyroclastic blobs that have a large range of sizes. The smaller clumps dissipate instantly to create a haze of low density cool gas whereas larger globules of >$10^3$ km size will retain their core integrity out to circumbinary distances that should explain X-ray emission outside of the Fe II obscuring envelope. The exact size distribution function of the clumps determines the X-ray characteristics of the nova. Finally, if the He/N emission is indeed due to the emitting ejecta clumps, whose initial sound speeds should be similar to the WD escape velocity (from virial arguments), roughly half of the line widths are attributable to the expansion of the individual clumps. Thus, actual blob ejection velocities from the WD are likely to be only ~50% of the velocity deduced from the line widths.

A schematic representation of the model proposed here is shown in Fig. 5. The visible He/N and hard X-ray emission, both lines and continuum, are formed predominantly in ejected globules within a region surrounding the WD. Depending on the mass ejection rate from the WD and the size distribution of the clumps the extent of the He/N emitting region might possibly exceed the size of the WD Roche lobe or it might be not much greater than the size of the WD. The Fe II gas component, which we believe originates in mass lost from the secondary star and is represented only in schematic form in the figure, is circumbinary in nature and generally envelopes the He/N radiating clumps. The predominant visible spectral class of a nova is therefore determined by the relative strengths of mass loss from the secondary and from the WD, and it can change between He/N and Fe II phases depending upon competition between continuing WD surface nuclear reaction driven mass loss and non-equilibrium responses to the outburst by the secondary star that no doubt drive its mass ejection. It is this variability that resulted in the visible spectrum of T Pyxidis evolving from the He/N phase to an Fe II phase and then back again to the He/N spectrum.

The line widths of the He/N spectra, which are determined by the ejection velocities of the globules *plus* the expansion velocities of each clump, are expected to be larger than the Fe II line widths, whose velocities are characterized by the lower ejection velocities of gas lost by the secondary star. The emission line widths of the final nebular stage will take on values between the line widths of the He/N and Fe II phases, depending on which of those components of gas dominates the nebular spectrum. Such behavior is just what was measured by Ederoclite et al. (2013) for the T Pyxidis Hβ line widths, which were initially quite broad and then decreased as the He/N to Fe II spectral transition occurred. The line widths then increased again as the spectrum evolved through the second He/N phase and the nebular spectrum.

The He/N ejecta blobs interact with the circumbinary Fe II gas as they pass through it, so the two components do mix with each other. Subsequent emission, as in the nebular spectral phase, consists of contributions from both components. If one attempts to determine the individual composition of the two components of gas it is best done early in the decline before much mixing takes place.

The model we propose here is not a radical departure from the conventional picture that ejecta originate entirely from the WD and that inhomogeneities are present. Its significant features are that (a) a very wide range of ionization conditions exist within individual blobs of ejected gas that naturally account for the large differences observed simultaneously in X-ray and visible spectra, (b) the secondary star is normally a major source of ejecta, and (c) the primary emitting region of WD ejecta is within the WD Roche lobe. These last two features admittedly arise primarily from the interpretation of the spectra of just two objects, T Pyxidis and U Scorpii. The hybrid evolution of T Pyx from He/N to Fe II and then back to He/N class points to obscuration of the He/N emission by a separate, larger Fe II component of gas. The postoutburst X-ray spectrum of U Sco during eclipse indicates a relatively confined X-ray emitting region. The broader applicability of these observations to most novae does need to be established, but it does serve to explain naturally the rapid flux variations that are observed. The essential point is that *the physics inside rapidly expanding gaseous globules are central to explaining the spectral evolution of novae.*

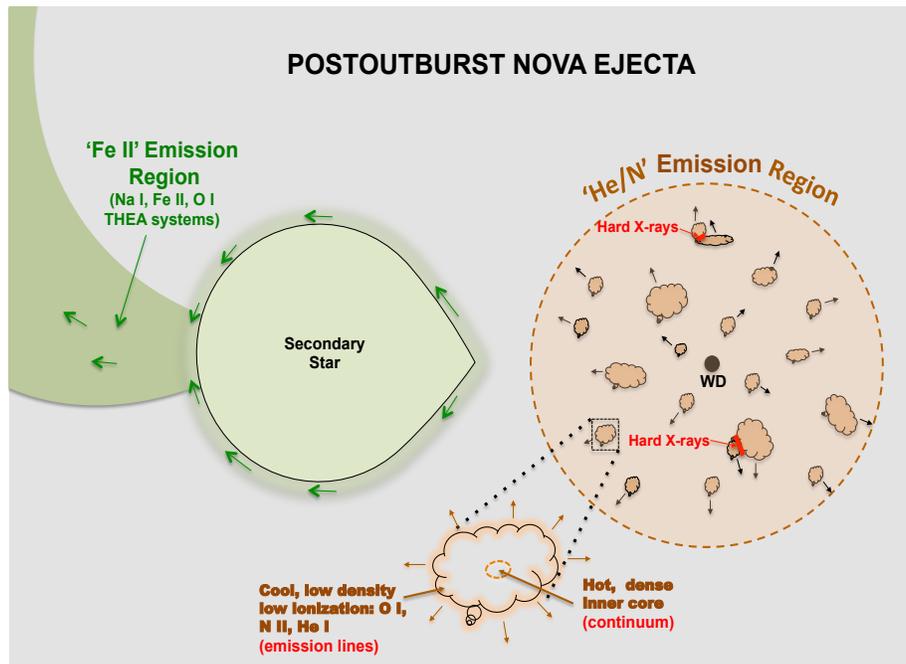

Fig. 5 – Discrete blob model for postoutburst novae that explains the main features of the optical and X-ray spectra.

    The spatially resolved nova shells of GK Per, HR Del, and T Pyx, which consist of many small clumps, are what the He/N WD ejecta should look like long after the outburst. The fact that strong inhomogeneities persist many decades after outburst is consistent with the existence of rather large dense globules of ejecta at the time of the outburst. The fact that there is presumably a more homogeneous stream of gas associated with the secondary star could explain the smoother ejected shell structure of some novae. For the sake of generality we have here ignored the complex geometry, sometimes conical and bipolar and with jets, that are known to occur from analyses of novae ejecta (Ribeiro et al. 2009; Chesneau et al. 2011, 2012). The conical and bipolar geometry can be accounted for by strong magnetic fields and by blockage of the ejecta from the accretion disk (Drake & Orlando 2010).

    Continuing coordinated spectroscopic follow up of novae over all wavelength regimes will eventually provide more detailed observations that enable the geometry of the postoutburst structure of novae to be refined. Since important aspects of the ejecta activity occur at the time of outburst spectral studies are best conducted as early as possible. Realistically, a more complete understanding of the outburst geometry may have to await LSST with its timely discovery of outbursts before maximum light occurs. It may be that an inhomogeneous and variable wind from the WD can explain many facets of the spectral development of novae in the X-ray, UV, visible, IR, and radio regions, however the attributes of expanding dense blobs of gas ejected from the WD that compete with mass loss from the secondary star is a model that accounts for many observed characteristics of novae and merits further study with detailed calculations.

    The author is grateful to Dr. V. Dogra for her discussions and results, and to my colleagues E. Mason, A. Ederoclite, and A. Bianchini for their continuing collaboration on studies of the spectral evolution of novae. The comments of the referee, Jay Gallagher, were especially appreciated in identifying areas that benefitted from more discussion and clarification.